\documentclass{article}
\usepackage{epsfig}
\usepackage{color}
\usepackage{amsmath}
\usepackage{amssymb}
\usepackage{lscape}
\usepackage{cite}

\makeatletter
\@addtoreset{equation}{section}

\begin{document}

\begin{flushright}
June 2017

KEK-TH-1985
%OCU-PHYS 449
\end{flushright}

\begin{center}

\vspace{3cm}

{\LARGE 
\begin{center}
Strong Coupling Limit

of 

A Family of Chern-Simons-matter Theories
\end{center}
}

\vspace{2cm}

Takao Suyama \footnote{e-mail address: tsuyama@post.kek.jp}

\vspace{1cm}

{\it 
KEK Theory Center, High Energy Accelerator Research Organization (KEK), 

Oho 1-1, Tsukuba, Ibaraki 305-0801, Japan
}

\vspace{2cm}

{\bf Abstract} 

\end{center}

We investigate the strong coupling limit of a family of Chern-Simons-matter theories in the planar limit. 
The family consists of ${\cal N}=3$ theories with the gauge group ${\rm U}(N_1)_{k_1}\times{\rm U}(N_2)_{k_2}$ coupled to $n$ bi-fundamental hypermultiplets. 
All observables which can be determined from the planar resolvent turn out to have finite limits in the large 't~Hooft coupling limit. 
Possible gravity duals are briefly discussed. 
We observe that Kac-Moody algebras govern the structure of the planar spectral curves of the theories. 

\newpage

\vspace{1cm}

\section{Introduction}

\vspace{5mm}

ABJM theory \cite{Aharony:2008ug} provides a prototypical example of AdS$_4$/CFT$_3$ correspondence. 
It is an ${\cal N}=6$ Chern-Simons-matter theory, and its gravity dual is M-theory on AdS$_4\times S^7/\mathbb{Z}_k$. 
It is natural to expect that this correspondence could be generalized by replacing the internal manifold $S^7/\mathbb{Z}_k$ in the gravity side with a less symmetric seven-dimensional manifold $B$. 
One should choose a manifold $B$ with some good properties in order to be able to perform the actual analysis. 
At the same time, one would like to choose $B$ from a wide range of possible manifolds so that the insight into AdS/CFT  correspondence \cite{Maldacena:1997re} can be gained by examining various examples. 

One such criterion for choosing $B$ is to require that the corresponding CFT$_3$ possesses ${\cal N}=3$ supersymmetry. 
Indeed, ${\cal N}=3$ supersymmetry for boundary theory is powerful enough to control quantum corrections, and also it is flexible enough to allow the construction of theories with various gauge groups and matter representations \cite{Gaiotto:2007qi}. 
A more detailed characterization of $B$ is given as follows. 
It is known that M-theory on the background of the form $\mathbb{R}^{1,2}\times M$ preserves ${\cal N}=3$ supersymmetry in three dimensions if and only if $M$ is hyper-K\"ahler. 
Suppose that $M$ has a conical singularity, and the base of the cone is $B$. 
One may put M2-branes to probe the singularity without breaking any supersymmetry, as long as the orientation of the M2-branes is appropriately chosen \cite{Gauntlett:1997pk}. 
The near-horizon geometry of the M2-branes is then AdS$_4\times B$ which is expected to be the dual gravity background for the worldvolume theory of the M2-branes. 
Therefore, one should choose $B$ to be a 3-Sasakian manifold (see e.g. \cite{Sparks:2010sn}). 

Suppose that the worldvolume theory of the M2-branes in the background discussed above is an ${\cal N}=3$ Chern-Simons-matter theory. 
Then, the hyper-K\"ahler manifold $M$ should be obtained as the moduli space of vacua of this theory. 
The moduli spaces were investigated in \cite{Imamura:2008nn} for ${\cal N}=3$ Chern-Simons-matter theories whose matter representation is specified by a circular quiver diagram. 
In \cite{Jafferis:2008qz}, this analysis was extended to theories corresponding to more general quiver diagrams, and it was shown that the moduli space is hyper-K\"ahler. 
It turned out that the dimension of the moduli space of a given theory is determined by the number of loops in the quiver diagram, whose explicit formula depends on the Chern-Simons levels. 
According to the formula in \cite{Jafferis:2008qz}, theories with A-type quiver diagrams have eight-dimensional moduli spaces, as was shown explicitly in \cite{Imamura:2008nn}, while those with DE-type quiver diagrams would have four-dimensional ones. 
Later, the analysis in \cite{Jafferis:2008qz} was generalized in \cite{Crichigno:2017rqg}, in order to treat non-toric hyper-K\"ahler manifolds as well, which showed that certain theories with D-type quivers also have eight-dimensional moduli spaces. 

A check of the correspondence can be performed by calculating the free energy. 
It was shown in \cite{Drukker:2010nc} that the free energy of ABJM theory behaves as 
\begin{equation}
F\ \sim\ \frac{\sqrt{2}}{3}\sqrt{k}N^{\frac32}
\end{equation}
in the large $N$ limit, which exactly reproduces the corresponding gravity result. 
This analysis can be extended to ${\cal N}=3$ cases. 
From the gravity side, the leading large $N$ behavior of the free energy is expected to be of the form \cite{Herzog:2010hf}
\begin{equation}
F\ \sim\ N^\frac32\sqrt{\frac{2\pi^6}{27{\rm vol}(B)}}
   \label{free energy}
\end{equation}
where ${\rm vol}(B)$ is the volume of $B$. 
The formula for ${\rm vol}(B)$ for a toric $B$ was obtained in \cite{Yee:2006ba}. 
This volume formula was extended in \cite{Crichigno:2017rqg} to non-toric ones. 
The large $N$ behavior (\ref{free energy}) of the free energy was reproduced from the corresponding ${\cal N}=3$ Chern-Simons-matter theories exactly by employing the technique developed in \cite{Herzog:2010hf}\footnote{
A similar technique was used in \cite{Suyama:2009pd} in the 't~Hooft limit. 
} based on the supersymmetric localization \cite{Kapustin:2009kz}. 
The technique was applied to various theories in \cite{Gulotta:2011si,Gulotta:2011aa,Crichigno:2012sk}. 
Similar analyses can be performed for ${\cal N}=2$ Chern-Simons-matter theories as well  \cite{Martelli:2011qj,Cheon:2011vi,Jafferis:2011zi,Martelli:2011fu,Amariti:2011uw,Gang:2011jj,Amariti:2012tj,Lee:2014rca}. 

\vspace{5mm}

Interestingly, there is one example of ${\cal N}=3$ duals which does not seem to have been investigated in detail, compared with the other examples. 
The M-theory background of this case consists of a 3-Sasakian manifold known as $N^{0,1,0}$ which is a coset ${\rm SU}(3)/{\rm U}(1)$. 
The cone over $N^{0,1,0}$ is $T^*\mathbb{P}^2$ which is a quite simple eight-dimensional hyper-K\"ahler manifold. 

The dual CFT$_3$ was proposed in \cite{Gukov:1999ya} to be a three-dimensional gauge theory with Chern-Simons terms where the gauge group is ${\rm U}(N)\times{\rm U}(N)$ coupled to {\it three} bi-fundamental hypermultiplets. 
The Chern-Simons levels were expected to be determined by a flux in the M-theory background, but their values were not specified explicitly. 
This proposal was further elaborated in \cite{Billo:2000zr} by showing the correspondence of BPS operators in the boundary theory with the Kaluza-Klein spectrum of the M-theory on AdS$_4\times N^{0,1,0}$. 
According to \cite{Jafferis:2008qz}, the Chern-Simons levels must satisfy $k_1+k_2\ne0$ in order to obtain $T^*\mathbb{P}^2$ as the moduli space. 
Recently, another dual CFT$_3$ was proposed in \cite{Gaiotto:2009tk}. 
The proposed theory is a deformation of ABJM theory by adding fundamental hypermultiplets. 
It was shown that quantum corrections to the classical moduli space is crucial for this duality. 
This proposal was supported by calculating the superconformal index which matches with the corresponding index in M-theory \cite{Cheon:2011th}. 

\vspace{5mm}

With these developments in mind, in this paper, we investigate ${\cal N}=3$ Chern-Simons-matter theories with the gauge group ${\rm U}(N_1)_{k_1}\times{\rm U}(N_2)_{k_2}$ coupled to $n\ge3$ bi-fundamental hypermultiplets in the planar limit. 
Note that such a family of theories with $n=2$ includes ABJM theory. 
Another family with $n=1$ was recently discussed in \cite{Nosaka:2017ohr}. 
In our previous paper \cite{Suyama:2016nap}, we analyzed the saddle point equations for the localized partition function, and obtained a closed formula for the resolvent in terms of the theta functions. 
From the resolvent, the free energy and the vevs of BPS Wilson loops, which are of the same kind investigated in \cite{Drukker:2008zx,Chen:2008bp,Rey:2008bh} for ABJM theory, can be calculated as functions of the 't~Hooft couplings. 
We investigate the behavior of these observables in the large 't~Hooft coupling limit. 
Interestingly, we find that the behavior is quite different from the one observed in ABJM theory. 
It turns out that the free energy scales as $N^2$, just like the ordinary gauge theories, and the vevs of BPS Wilson loops approach constant values in the limit. 
In addition, other observables which can be determined from the planar resolvent are found to exhibit similar behaviors. 
This difference from the case of ABJM theory is rather surprising since the difference at the level of Lagrangians is just to increase the number of the hypermultiplets. 
Curiously, a difference can be seen also at the level of spectral curves. 

This paper is organized as follows. 
In section \ref{section-planar}, we show that the above-mentioned results for the Chern-Simons-matter theories are derived from the resolvent obtained in \cite{Suyama:2016nap}. 
Possible gravity duals for the theories are briefly reconsidered  in section \ref{section-dual}, taking into account the results obtained in the previous section. 
In section \ref{section-curve}, we observe that differences among the ${\cal N}=3$ Chern-Simons-matter theories investigated in \cite{Jafferis:2008qz} can be seen in their spectral curves which are governed by associated Kac-Moody algebras. 
Section \ref{section-discuss} is devoted to discussion. 
Some details on our resolvent are reviewed in Appendix \ref{details}. 
Appendix \ref{infinite sheets} contains a proof for a fact on the spectral curve. 

\vspace{1cm}

\section{Planar results on Chern-Simons-matter theories} \label{section-planar}

\vspace{5mm}

In this section, we investigate theories in a family CSM$(n)$ of ${\cal N}=3$ Chern-Simons-matter theories for a positive integer $n$ by using the corresponding matrix models. 
Each theory in CSM$(n)$ has a gauge group of the form ${\rm U}(N_1)_{k_1}\times{\rm U}(N_2)_{k_2}$ and $n$ 
hypermultiplets in the bi-fundamental representation of the gauge group. 
The action of the theory is completely specified by these data due to ${\cal N}=3$ supersymmetry \cite{Gaiotto:2007qi}. 
ABJM theory \cite{Aharony:2008ug}\cite{Aharony:2008gk} and GT theory \cite{Gaiotto:2009mv} are members of CSM$(2)$. 
We are mainly interested in the theories in CSM$(3)$ since there exists a proposal for a gravity dual \cite{Gukov:1999ya}\cite{Billo:2000zr}. 

\vspace{5mm}

\subsection{Planar resolvent}

\vspace{5mm}

The partition function of a theory in CSM$(n)$ defined on $S^3$ can be given in terms of a finite-dimensional integral \cite{Kapustin:2009kz} as 
\begin{equation}
Z\ =\ \int d^{N_1}u\,d^{N_2}w\,\exp\left[ \frac{ik_1}{4\pi}\sum_{i=1}^{N_1}(u_i)^2+\frac{ik_2}{4\pi}\sum_{a=1}^{N_2}(w_a)^2 \right]\frac{\prod_{i<j}^{N_1}(\sinh\frac{u_i-u_j}2)^2\prod_{a<b}^{N_2}(\sinh\frac{w_a-w_b}2)^2}{\prod_{i=1}^{N_1}\prod_{a=1}^{N_2}(\cosh\frac{u_i-w_a}2)^n}. 
   \label{Z}
\end{equation}
This integral can be regarded as the partition function of a matrix model. 
As usual for matrix models, we may take the planar limit which makes the saddle point approximation for the integral (\ref{Z}) exact. 
The planar limit is defined by introducing an auxiliary parameter $k$ as 
\begin{equation}
k\ \to\ \infty \hspace{5mm} \mbox{with } \hspace{5mm} t_\alpha\ :=\ \frac{2\pi iN_\alpha}{k}, \hspace{5mm} \kappa_\alpha\ :=\ \frac{k_\alpha}{k} \hspace{5mm} \mbox{fixed.} \hspace{5mm} (\alpha=1,2)
   \label{planar limit}
\end{equation}
In this limit, the integral in (\ref{Z}) is dominated by the saddle point which satisfies the following equations: 
\begin{eqnarray}
\frac{k_1}{2\pi i}u_i &=& \sum_{j\ne i}^{N_1}\coth\frac{u_i-u_j}2-\frac n2\sum_{a_1}^{N_2}\tanh\frac{u_i-w_a}2,      
   \label{original saddle1} \\
\frac{k_2}{2\pi i}w_a &=& \sum_{b\ne a}^{N_2}\coth\frac{w_a-w_b}2-\frac n2\sum_{i=1}^{N_1}\tanh\frac{w_a-u_i}2. 
   \label{original saddle2}
\end{eqnarray}

Let $\{\bar{u}_i, \bar{w}_a\}$ denote the solution of these equations. 
Because of the presence of $i$ in the left-hand side, $\bar{u}_i$ and $\bar{w}_a$ are complex numbers in general. 
For later convenience, the eigenvalues $\bar{u}_i$ and $\bar{w}_a$ are labeled such that 
\begin{equation}
{\rm Re}(\bar{u}_i)\ \le \ {\rm Re}(\bar{u}_{i+1}), \hspace{1cm} {\rm Re}(\bar{w}_a)\ \le\ {\rm Re}(\bar{w}_{a+1})
\end{equation}
holds. 
We assume that both $\bar{u}_i$ and $\bar{w}_a$ approach zero in the small 't~Hooft coupling limit, since in this limit the left-hand sides dominate in the equations. 

Some physical observables can be calculated in the planar limit. 
The free energy is obtained from the saddle point value of the integral (\ref{Z}). 
In addition, there are two BPS Wilson loops for ${\rm U}(N_1)$ gauge fields and ${\rm U}(N_2)$ gauge fields. 
Their vevs in the planar limit are given as 
\begin{equation}
\langle W_1\rangle\ =\ \frac1{N_1}\sum_{i=1}^{N_1}e^{\bar{u}_i}, \hspace{1cm} \langle W_2\rangle\ =\ \frac1{N_2}\sum_{a=1}^{N_2}e^{\bar{w}_a}. 
\end{equation}

\vspace{5mm}

The information of the theory in the planar limit is encoded in a row-vector-valued resolvent \cite{Suyama:2016nap}
\begin{equation}
v(z)\ :=\ (v_1(z),v_2(z)). 
\end{equation}
The components $v_\alpha(z)$ are defined as 
\begin{equation}
v_1(z)\ :=\ \lim \frac{t_1}{N_1}\sum_{i=1}^{N_1}\frac{z+e^{\bar{u}_i}}{z-e^{\bar{u}_i}}, \hspace{1cm} v_2(z)\ :=\ \lim \frac{t_2}{N_2}\sum_{a=1}^{N_2}\frac{z-e^{\bar{w}_a}}{z+e^{\bar{w}_a}}, 
\end{equation}
where $\lim$ indicates the planar limit (\ref{planar limit}). 
The function $v_1(z)$ has a square-root branch cut on an interval $I_1$ in $\mathbb{C}$, and holomorphic elsewhere, including the point at infinity. 
The branch points $a_1$ and $b_1$ on $I_1$ are given as 
\begin{equation}
a_1\ :=\ e^{\bar{u}_1}, \hspace{1cm} b_1\ :=\ e^{\bar{u}_{N_1}}. 
   \label{def of a1}
\end{equation}
Due to the symmetry of the equations (\ref{original saddle1})(\ref{original saddle2}), they satisfy 
\begin{equation}
a_1b_1\ =\ 1. 
   \label{ab1}
\end{equation}
The other function $v_2(z)$ has similar properties. 
The branch points $a_2$ and $b_2$ are given as 
\begin{equation}
a_2\ :=\ -e^{\bar{w}_1}, \hspace{1cm} b_2\ :=\ -e^{\bar{w}_{N_2}}
   \label{def of a2}
\end{equation}
which also satisfy 
\begin{equation}
a_2b_2\ =\ 1. 
   \label{ab2}
\end{equation}
Note that $|a_\alpha|\le1$ are satisfied. 

The resolvent $v(z)$ is determined by specifying two complex parameters, say $a_\alpha$. 
These two parameters are related to the values of the 't~Hooft couplings $t_\alpha$ for given $\kappa_\alpha$ via 
\begin{equation}
v(0)\ =\ -(t_1,t_2). 
   \label{tHooft0}
\end{equation}

\vspace{5mm}

It was found in \cite{Suyama:2016nap} that it is convenient to investigate a derivative $zv'(z)$ of the resolvent, instead of $v(z)$ itself, since the former can be obtained explicitly. 
In the following, we assume $n\ne2$. 

Let $c:=(c_1,c_2)$ be the row vector satisfying 
\begin{equation}
(2\kappa_1,2\kappa_2)\ =\ (c_1,c_2)\left[
\begin{array}{cc}
2 & -n \\
-n & 2
\end{array}
\right]. 
   \label{vector c}
\end{equation}
Note that the solution of this equation exists for $n\ne2$. 
Let us introduce another row-vector-valued function $f(z)$ such that $zv'(z)$ can be written as 
\begin{equation}
zv'(z)\ =\ c+\frac1{s(z)}f(z), \hspace{1cm} s(z)\ :=\ \sqrt{(z-a_1)(z-b_1)(z-a_2)(z-b_2)}. 
   \label{v to f}
\end{equation}
In order to determine the explicit form of $f(z)$, we introduce a new variable $u$  defined as 
\begin{equation}
u(z)\ :=\ \frac{\varphi(z)}{2\varphi(b_1)}, \hspace{1cm} \varphi(z)\ :=\ \int_{a_1}^z\frac{d\xi}{s(\xi)}, 
\end{equation}
where the integration contour for $\varphi(b_1)$ lies above the segment $I_1$. 
The saddle point equations (\ref{original saddle1})(\ref{original saddle2}) imply that, as a function of $u$, $f(z)$ can be written as 
\begin{equation}
f(z(u))\ =\ \left( \tilde{f}(u), \tilde{f}(-u) \right)S^{-1}, \hspace{1cm} S\ :=\ \left[  
\begin{array}{cc}
1 & 1 \\
-e^{\pi i\nu} & -e^{-\pi i\nu}
\end{array}
\right]
   \label{S}
\end{equation}
where $\nu$ is related to $n$ by $n=2\cos\pi\nu$. 
The scalar function $\tilde{f}(u)$ is required to satisfy 
\begin{equation}
\tilde{f}(u+1)\ =\ \tilde{f}(u), \hspace{1cm} \tilde{f}(u+\tau)\ =\ e^{2\pi i\nu}\tilde{f}(u)
   \label{conditions tilde{f}}
\end{equation}
where $\tau:=2u(a_2)$. 
Therefore, $\tilde{f}(u)$ can be written in terms of the theta functions. 
The explicit form of $\tilde{f}(u)$ is given in Appendix \ref{details}. 

Since what we have obtained is $zv'(z)$, the relation (\ref{tHooft0}) cannot be used to determine $t_\alpha$ for a given $a_\alpha$. 
An alternative way to recover $t_\alpha$ is to use the formula 
\begin{equation}
t_\alpha\ =\ -\frac12\int_{C_\alpha}\frac{dz}{2\pi i}\frac{\log z}zzv_\alpha'(z), 
   \label{tHooft}
\end{equation}
where $C_1$ ($C_2$) is a contour encircling the segment $I_1$ ($I_2$) counterclockwise. 

In addition to $t_\alpha$, the expansion of $zv'(z)$ provides the vevs $\langle W_\alpha\rangle$ of BPS Wilson loops as 
\begin{equation}
zv'(z)\ =\ -2\left( t_1\langle W_1\rangle, -t_2\langle W_2\rangle \right)z+O(z^2). 
   \label{WL}
\end{equation}
Combining (\ref{tHooft}) and (\ref{WL}), the vevs $\langle W_\alpha\rangle$ can be given as functions of $t_\alpha$. 

\vspace{5mm}

\subsection{Small 't~Hooft coupling limit}

\vspace{5mm}

Various quantities can be calculated perturbatively when the 't~Hooft couplings $t_\alpha$ are small. 
For example, the vevs $\langle W_\alpha\rangle$ are given as \cite{Suyama:2016nap} 
\begin{eqnarray}
\langle W_1\rangle 
&=& 1+\frac{t_1}{2\kappa_1}+\frac16\left( \frac{t_1}{\kappa_1} \right)^2\left( 1-\frac1{4N_1^2} \right)-\frac n8\frac{t_1t_2}{\kappa_1^2}+O(t^3), \\
\langle W_2\rangle 
&=& 1+\frac{t_2}{2\kappa_2}+\frac16\left( \frac{t_2}{\kappa_2} \right)^2\left( 1-\frac1{4N_2^2} \right)-\frac n8\frac{t_1t_2}{\kappa_2^2}+O(t^3), 
\end{eqnarray}
which can be obtained by using the method developed in \cite{Kapustin:2009kz}. 
This result was reproduced in \cite{Suyama:2016nap} from the resolvent $v(z)$ reviewed above by using the formulas (\ref{tHooft})(\ref{WL}) for cases when 
$N_1=N_2$ and $k_1=\pm k_2$ are satisfied. 
This can be regarded as a non-trivial check of the validity of the resolvent $v(z)$ obtained in \cite{Suyama:2016nap}. 

\vspace{5mm}

The method in \cite{Kapustin:2009kz}, however, does not give us any information on the configuration of the eigenvalues $\{\bar{u}_i,\bar{w}_a\}$, and therefore, the positions $a_\alpha$ of the branch points of $v(z)$. 
To obtain such information, it is better to solve the saddle point equations (\ref{original saddle1})(\ref{original saddle2}) perturbatively, as in \cite{Suyama:2009pd}. 

In the following, we will restrict ourselves to the case $N_1=N_2=:N$ for simplicity. 
This implies that $t_1=t_2=:t$ holds. 
The analysis of this case should be the first step toward the understanding of the general case. 

Recall that the saddle point equations in this case are 
\begin{eqnarray}
\frac{k_1}{2\pi i}u_i &=& \sum_{j\ne i}^{N}\coth\frac{u_i-u_j}2-\frac n2\sum_{a=1}^{N}\tanh\frac{u_i-w_a}2, 
   \label{saddle1} \\
\frac{k_2}{2\pi i}w_a &=& \sum_{b\ne a}^{N}\coth\frac{w_a-w_b}2-\frac n2\sum_{i=1}^{N}\tanh\frac{w_a-u_i}2. 
   \label{saddle2}
\end{eqnarray}

Assume that $t$ is small, and introduce rescaled variables $x_i$ and $y_a$ defined such that 
\begin{equation}
u_i\ =\ \sqrt{t}\,x_i, \hspace{1cm} w_a\ =\ \sqrt{t}\,y_a 
\end{equation}
holds. 
Then, the saddle point equations can be expanded in $t$. 
The leading order terms give the following simple equations: 
\begin{equation}
\kappa_1 x_i\ =\ \frac2N\sum_{j\ne i}^N\frac1{x_i-x_j}, \hspace{1cm} \kappa_2y_a\ =\ \frac2N\sum_{b\ne a}\frac1{y_a-y_b}. 
\end{equation}
Each of these equations are the same as the saddle point equations of the Gaussian matrix model. 
The planar solution of them is encoded in 
\begin{eqnarray}
\omega_1(z)\ :=\ \lim\frac1N\sum_{i=1}^N\frac1{z-x_i}\ =\ \frac{\kappa_1}2\left[ z-\sqrt{z^2-\frac{4}{\kappa_1}}\ \right], \\
\omega_2(z)\ :=\ \lim\frac1N\sum_{a=1}^N\frac1{z-y_a}\ =\ \frac{\kappa_2}2\left[ z-\sqrt{z^2-\frac{4}{\kappa_2}}\ \right]. 
\end{eqnarray}
The perturbative corrections to this solution can be systematically calculated, as shown in \cite{Suyama:2009pd} for ABJM theory. 

The discontinuity of $\omega_1(z)$ and $\omega_2(z)$ encodes the distributions of $\{x_i\}$ and $\{y_a\}$ which then give the distributions of $\{u_i\}$ and $\{w_a\}$.  
The definitions (\ref{def of a1})(\ref{def of a2}) of the parameters $a_1$ and $a_2$ imply that they are given as 
\begin{equation}
a_1\ \sim\ \exp\left( -2\sqrt{\frac{t}{\kappa_1}}\ \right), \hspace{1cm} a_2\ \sim\ -\exp\left( -2\sqrt{\frac{t}{\kappa_2}}\ \right)
   \label{weak coupling}
\end{equation}
for small $t$. 

One may notice that the relation between $a_1$ and $a_2$ becomes quite simple when $\kappa_1=\kappa_2$ holds. 
In this case, the parameters are related simply as 
\begin{equation}
a_1\ =\ -a_2\ =:\ a. 
   \label{a1=-a2}
\end{equation}
This relation is satisfied when the equalities $\bar{u}_i=\bar{w}_i$ hold. 
Since this equality is compatible with the full saddle point equations (\ref{saddle1})(\ref{saddle2}) with $k_1=k_2$, the relation (\ref{a1=-a2}) should hold beyond perturbation. 
Therefore, any observables which can be derived from the resolvent $v(z)$ are functions of $a$. 
Without loss of generality, we can choose $\kappa_1=\kappa_2=1$. 

Note that the explicit form of $zv'(z)$ shown in Appendix \ref{details} is written in terms of the parameters $u_0, u_\infty$, and $\tau$. 
The definition of $u(z)$ implies 
\begin{equation}
u_\infty\ =\ u_0-\frac12. 
   \label{u02u00}
\end{equation}
When the equality $a_1=-a_2$ holds, $u_0$ is given by $\tau$ as 
\begin{equation}
u_0\ =\ \frac14\tau. 
\end{equation}
Therefore, all the observables are also functions of $\tau$. 
The relation between $a$ and $\tau$ can be obtained from the inverse of $u(z)$ given as 
\begin{equation}
z(u)\ =\ -\frac{\vartheta_1(u-u_0)\vartheta_1(u+u_0)}{\vartheta_1(u-u_\infty)\vartheta_1(u+u_\infty)}. 
   \label{z(u)}
\end{equation}
Since $z=a$ corresponds to $u=0$ by definition, $a$ is related to $\tau$ as 
\begin{equation}
a\ =\ -\left( \frac{\vartheta_1(\frac14\tau)}{\vartheta_1(\frac14\tau-\frac12)} \right)^2. 
   \label{a2tau}
\end{equation}
In the following, we choose $\tau$ as the parameter specifying the resolvent. 

It is important to notice that not all values of $\tau$ are physically relevant. 
Since the 't~Hooft couplings are defined as (\ref{planar limit}), their physical values are purely imaginary. 
Therefore, $\tau$ must be chosen such that the integrals (\ref{tHooft}) take purely imaginary values. 
The physical values of $\tau$ form a curve $\gamma$ in the complex $\tau$-plane. 

The asymptotic behavior of $\gamma$ in the small 't~Hooft coupling limit is determined as follows. 
When $a$ approaches 1, the definition of $\tau$ implies that ${\rm Im}(\tau)$ diverges to $+\infty$. 
In this limit, the relation (\ref{a2tau}) becomes 
\begin{equation}
a\ \sim\ 1-4e^{\frac12\pi i\tau}. 
\end{equation}
On the other hand, (\ref{weak coupling}) implies 
\begin{equation}
a\ \sim\ 1-2\sqrt{t}
\end{equation}
for small $t$. 
Therefore, in the small 't~Hooft coupling limit, the physical curve $\gamma$ approaches the line 
\begin{equation}
{\rm Re}(\tau)\ =\ \frac12. 
\end{equation}

\vspace{5mm}

\subsection{Large 't~Hooft coupling limit}

\vspace{5mm}

Our interest is in the large 't~Hooft coupling limit. 
The limit can be obtained by following the curve $\gamma$ in the direction opposite to the small 't~Hooft coupling limit discussed above. 
Recall that the curve $\gamma$ is defined as ${\rm Re}(t)=0$. 
Since $t$ is given in terms of the integral of a complicated function, it looks quite difficult to determine the curve $\gamma$ analytically. 
Instead, we evaluate the integral (\ref{tHooft}) numerically, and find out where ${\rm Re}(t)=0$ holds in the $\tau$-plane.  

\begin{figure}
\begin{center}
\includegraphics{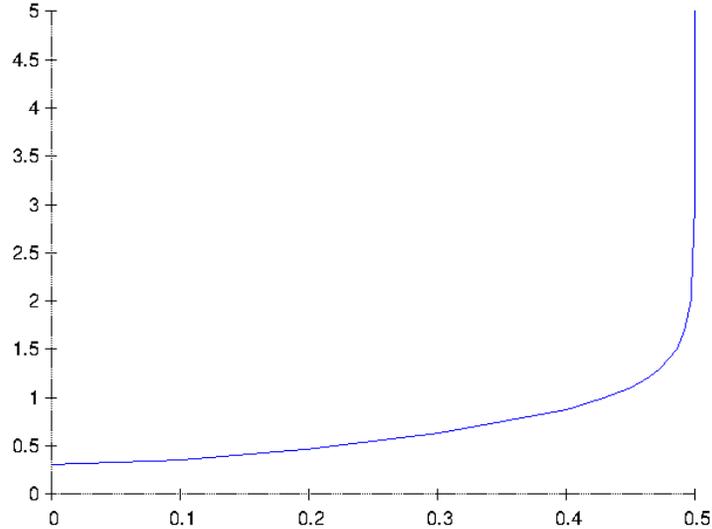}
\end{center}
\caption{The plot of the curve $\gamma$ for $n=3$ on which $t$ is purely imaginary. 
The curve $\gamma$ approaches the line ${\rm Re}(\tau)=\frac12$ in the limit $t\to0$. 
It terminates on the imaginary axis at which $t$ diverges. }
  \label{plot}
\end{figure}

The plot of the curve $\gamma$ for $n=3$ is shown in Figure \ref{plot}. 
The curve $\gamma$ terminates on the imaginary axis at which the 't~Hooft coupling $t$ diverges as 
\begin{equation}
t\ \sim\ \frac{r}{\tau-\nu}, \hspace{1cm} r\ =\ 1.27\,i,
   \label{pole}
\end{equation}
where $\nu=0.306\,i$ for $n=3$. 
This pole comes from the factor $(1+z(u_\nu))^{-1}$ in the function $g(u)$ defined in Appendix \ref{details}. 
The singularity at $\tau=\nu$ was pointed out in \cite{Suyama:2016nap}. 

The values of the vevs $\langle W_1\rangle$ and $\langle W_2\rangle$ are the same since $\bar{u}_i=\bar{w}_i$ holds. 
Their values in the large $t$ limit are 
\begin{eqnarray}
\langle W_1\rangle\ =\ \langle W_2\rangle 
&=& -\frac1{4\varphi(b)(1+e^{\pi i\nu})t}\left[ g'(u_0)G(u_0)+g(u_0)G'(u_0) \right] \nonumber \\
&\to& 5.77\ . 
\end{eqnarray}
Note that these are finite even in the large $t$ limit, in contrast to ABJM theory and GT theory. 
The finiteness can be anticipated from the eigenvalue distribution. 
Recall that the large $t$ limit corresponds to the limit $\tau\to\nu$. 
The relation (\ref{a2tau}) implies that $a=1.19\times10^{-2}$ in the limit which is small but finite. 
Then, the value of $b=a^{-1}$ is also finite. 
The following inequality 
\begin{equation}
|\langle W_1\rangle|\ \le\ \frac1{N_1}\sum_{i=1}^{N_1}e^{{\rm Re}(\bar{u}_i)}\ \le\ |b|
\end{equation}
implies that $\langle W_1\rangle$ must be finite. 

\vspace{5mm}

In fact, it turns out that all the observables derived from the resolvent $v(z)$ are finite in the large $t$ limit. 
To show this, consider a general situation in which $\rho(x;t)$ is a density function with a parameter $t$  whose support is an interval $I$. 
The expectation value of ${\cal O}(x)$ is defined as 
\begin{equation}
\langle {\cal O}\rangle_t\ :=\ \int_I dx\,\rho(x;t){\cal O}(x). 
   \label{<O>}
\end{equation}
Define a resolvent 
\begin{equation}
\omega(z;t)\ :=\ t\int_Idx\,\rho(x;t)\frac{z+x}{z-x}. 
\end{equation}
The expectation value (\ref{<O>}) can be written in terms of $\omega(z;t)$ as 
\begin{equation}
\langle O\rangle_t\ =\ \frac1{2t}\int_C\frac{dz}{2\pi i}{\cal O}(z)\frac{\omega(z;t)}{z}, 
\end{equation}
where $C$ is a contour in the $z$-plane encircling $I$ counterclockwise. 
Suppose that a function ${\cal F}(x)$ satisfies 
\begin{equation}
{\cal F}'(x)\ =\ \frac1x{\cal O}(x). 
\end{equation}
Then, $\langle{\cal O}\rangle_t$ can be written as 
\begin{equation}
\langle{\cal O}\rangle_t\ =\ -\frac1{2t}\int_C\frac{dz}{2\pi i}\frac{{\cal F}(z)}{z}\cdot z\omega'(z;t). 
\end{equation}
This integral has a finite value in the limit $t\to\infty$ if $t^{-1}z\omega'(z;t)$ has a finite limit at any points on the contour $C$. 

In the case of the Chern-Simons-matter matrix models, the condition for the finiteness turns out to be 
\begin{equation}
\lim_{t\to +\infty i}\left| t^{-1}\tilde{f}(u) \right|\ <\ \infty. \hspace{1cm} \left( {\textstyle -\frac12\le u\le+\frac12} \right)
\end{equation}
This can be shown to be the case numerically. 
Since this implies that the eigenvalue distribution of this matrix model has a finite limit, all the quantities calculated by using this eigenvalue distribution must be finite. 
Recall that the planar free energy is equal to $N^2$ times a quantity calculated by the eigenvalue distribution. 
Therefore, it scales as $N^2$, as for usual gauge theories. 

\vspace{5mm}

If $n$ is large, then ${\rm Im}(\nu)$ becomes large. 
Then, since the pole of $t$ is located at the point $\tau=\nu$, the $q$-expansion of the theta functions can be used for the analysis of the large $t$ limit. 
The behavior of $t$ around $\tau=\nu$ turns out to be 
\begin{equation}
t\ \sim\ \frac r{\tau-\nu}, \hspace{1cm} r\ =\ \frac{2\,i}{\pi(e^{-\pi i\nu}-1)}. 
\end{equation}
This pole structure is qualitatively the same as (\ref{pole}) observed in the numerical result for $n=3$. 

The vevs $\langle W_\alpha\rangle$ in the large $t$ limit are 
\begin{equation}
\langle W_1\rangle\ =\ \langle W_2\rangle\ \to\ \frac1{1+e^{\pi i\nu}}. 
\end{equation}
This limiting value approaches 1 in the large $n$ limit. 
This can be anticipated from the saddle point equations (\ref{saddle1})(\ref{saddle2}). 
As ${\rm Im}(\nu)$ becomes large, the attractive forces among the eigenvalues due to the second terms in the right-hand side of (\ref{saddle1})(\ref{saddle2}) become strong. 
Then, the eigenvalue distribution shrinks to two points in the limit. 

\vspace{5mm}

\subsection{Small $\tau$ limit}

\vspace{5mm}

We observed in the previous subsection that the behavior of the vevs $\langle W_\alpha\rangle$ for the theories in CSM$(n\ge3)$ is quite different from that of ABJM theory and GT theory. 
From the matrix model point of view, the difference comes from the fact that, in the former theories, the extent of the eigenvalue distributions are kept finite in the large 't~Hooft coupling limit, while in the latter theories, the eigenvalue distributions grow indefinitely. 
It is curious to know what happens if the eigenvalue distributions grow indefinitely in the former theories. 
It is not a strong coupling limit of the Chern-Simons-matter theories, but it could be realized by an analytic continuation of the parameters. 

Since the branch points are constrained as (\ref{ab1})(\ref{ab2}), an infinitely long distribution of the eigenvalues can be realized only when the parameter $a(=a_1=-a_2)$ approaches zero. 
Small $a$ limit corresponds to a small $\tau$ limit. 
Due to the relation (\ref{a2tau}), they are related as  
\begin{equation}
-\log a\ \sim\ \frac{\pi i}{2\tau}. 
\end{equation}

It turns out that it is not straightforward to take the limit $\tau\to0$. 
Since the factor $(1+z(u_\nu))^{-1}$, which is the origin of the pole of $t$ at $\tau=\nu$, has in fact infinitely many poles on the $\tau$-plane for $n>2$. 
The positions of the poles are 
\begin{equation}
\tau\ =\ \frac\nu{2l+1}+\frac m{2l+1}. \hspace{1cm} (l,m\in\mathbb{Z})
\end{equation}
Therefore, the small $\tau$ limit taken along the imaginary axis, for example, is ill-defined. 

One way to avoid these singularities is to take the following limit: 
\begin{equation}
\tau\ =\ \varepsilon e^{i\theta}, \hspace{1cm} \varepsilon\ \to\ 0 
\end{equation}
with a fixed $\theta$. 
It turns out that if $\theta$ is in the range 
\begin{equation}
0\ <\ \cot\theta\ <\ \frac1{2\,{\rm Im}(\nu)}, 
\end{equation}
then all the poles can be avoided. 
At the same time, the expansion of the theta functions, after the transformation $\tau\to-\frac1\tau$, behaves well. 

The integral formula (\ref{tHooft}) gives the small $\tau$ behavior of $t$. 
The leading order term turns out to be 
\begin{equation}
t\ \sim\ \frac C\tau, \hspace{1cm} C\ :=\ \frac1{\sinh\left( \frac\pi2{\rm Im}(\nu) \right)}\sum_{n=1}^\infty\frac{(-i)^n}{n+\nu}. 
\end{equation}
Note that the sum in the coefficient $C$ for $n=3$ is finite: 
\begin{equation}
\sum_{n=1}^\infty\frac{(-i)^n}{n+\nu}\ =\ -0.591-0.641\,i. 
\end{equation}
Then, an upper bound of the vevs $\langle W_\alpha\rangle$ is obtained as 
\begin{equation}
|\langle W_\alpha\rangle|\ \le\ |b|\ \lesssim\ \left| \exp\left( \frac{\pi i}{2C}t \right) \right|. 
\end{equation}
As in the cases for ABJM theory and GT theory, this upper bound is actually saturated. 
The formula (\ref{WL}) implies 
\begin{equation}
\langle W_\alpha\rangle\ \sim\ \exp\left( \frac{\pi i}{2C}t \right)
\end{equation}
for large $t$. 
This behavior is different from the Wilson loops in ABJM theory and GT theory. 

\vspace{5mm}

\subsection{The case $k_1=-k_2$}

\vspace{5mm}

The weak coupling behavior (\ref{weak coupling}) indicates that there is another case $k_1=-k_2$ for which the analysis is rather simple. 
Assuming that $t$ takes a physical value, that is $t\in i\mathbb{R}$, the parameters $a_1$ and $a_2$ are related as 
\begin{equation}
a_1\ =\ -a_2^*. 
   \label{rel2}
\end{equation}
This is realized when the equalities $u_i=w_i^*$ are satisfied. 
This equality is compatible with the full saddle point equations (\ref{saddle1})(\ref{saddle2}), and therefore, the relation (\ref{rel2}) should hold beyond perturbation. 
Indeed, such eigenvalue distributions were discussed in \cite{Suyama:2009pd}\cite{Herzog:2010hf}. 

When the relation (\ref{rel2}) holds, the parameters $u_0,u_\infty$, and $\tau$ in the resolvent $v(z)$ are related among them. 
In addition to the relation (\ref{u02u00}), one can show that $\tau$ is given as 
\begin{equation}
\tau\ =\ 4i\,{\rm Im}(u_0). 
\end{equation}
Therefore, all the observables derived from the resolvent $v(z)$ are functions of $u_0$. 
Let $u_0$ be written as 
\begin{equation}
u_0\ =\ u_R+\frac14\tau, \hspace{1cm} u_R\ \in\ \mathbb{R}, 
\end{equation}
the parameter $a_1$ is given as 
\begin{equation}
a_1\ \sim\ 1-4e^{\frac12\pi i\tau}e^{2\pi iu_R}
\end{equation}
for large ${\rm Im}(\tau)$. 
The relation (\ref{weak coupling}) for small $t$ implies that the physical curve $\gamma$, now on the $u_0$-plane, approaches the line 
\begin{equation}
{\rm Re}(u_0)\ =\ \frac18
\end{equation}
in the small $t$ limit. 

The large $t$ limit can be analyzed in the same way as in the case $k_1=k_2$. 
The plot of the curve $\gamma$ for $n=3$ turns out to be quite similar to figure \ref{plot}. 
The curve $\gamma$ terminates on the imaginary axis of the $u_0$-plane. 
The intersection point is at 
\begin{equation}
u_0\ =\ \frac 14\nu, 
\end{equation}
where $t$ diverges. 
The vevs $\langle W_\alpha\rangle$ are finite even in the large $t$ limit. 

\vspace{1cm}

\section{A gravity dual revisited} \label{section-dual}

\vspace{5mm}

In this section, we briefly revisit the discussion on possible gravity duals of theories in ${\rm CSM}(n\ge3)$. 

The first step for the discussion is to examine the classical moduli space of vacua of the theory. 
The dimension $D$ of the moduli space for a given Abelian Chern-Simons-matter theory is given by the matter representation and the Chern-Simons levels $k_i$ \cite{Jafferis:2008qz}. 
Let $l$ be the number of loops in the quiver diagram dictating the matter representation of the theory. 
The dimension $D$ is given as 
\begin{equation}
D\ =\ \left\{
\begin{array}{cc}
4(l+1), & \left( \sum_ik_i=0 \right) \\ [2mm]
4l. & (\sum_ik_i\ne0)
\end{array}
\right.
\end{equation}
For example, ABJM theory corresponds to a quiver diagram with $l=1$ and the two Chern-Simons levels are $k$ and $-k$. 
Then, the above formula implies $D=8$. 
This coincides with the dimension of the moduli space $\mathbb{C}^4/\mathbb{Z}_k$. 
For the theory in ${\rm CSM}(n\ge3)$, the corresponding quiver diagram has $l=n-1$ loops. 
Therefore, the moduli space can be eight-dimensional only when $n=3$ and $k_1+k_2\ne0$ are satisfied. 
This is one reason that we focused our attention mainly on the theories with $n=3$ and $k_1=k_2$. 

In the following, we assume $n=3$ and $k_1+k_2\ne0$. 
The moduli space turns out to be $T^*\mathbb{P}^2$ \cite{Gukov:1999ya}. 
This is an eight-dimensional hyper-K\"ahler cone whose base manifold is $N^{0,1,0}={\rm SU}(3)/{\rm U}(1)$. 
Then, one possibility would be that the theories in ${\rm CSM}(3)$ would be dual to (suitable deformations of) M-theory on AdS$_4\times N^{0,1,0}$ \cite{Gukov:1999ya}\cite{Billo:2000zr}. 
Note that there could be quantum corrections to the moduli space, as shown in \cite{Gaiotto:2009tk} for a flavored ABJM theory. 

\vspace{5mm}

This proposed gravity dual is, however, rather different from the dual of ABJM theory, that is, M-theory on AdS$_4\times S^7/\mathbb{Z}_k$. 
In the latter, the planar limit corresponds to a Type IIA limit since the limit involves $k\to\infty$ in which a circle direction in $S^7$ shrinks. 
On the other hand, in the former, the planar limit seems to be still eleven-dimensional since $N^{0,1,0}$ is independent of the Chern-Simons levels. 

To understand the origin of the difference, it would be instructive to recall how the moduli space of vacua is determined for Abelian Chern-Simons-matter theories. 
The relevant part of the action is 
\begin{equation}
\frac{ik_1}{4\pi}\int A_1\wedge dA_1+\frac{ik_2}{4\pi}\int A_2\wedge dA_2-\int d^3x\,|\partial_\mu Y^I-i(A_1-A_2)_\mu Y^I|^2+\cdots. 
\end{equation}

If the Chern-Simons levels satisfy $k_1+k_2=0$, then by choosing a suitable linear combinations $A$ and $\tilde{A}$ of the gauge fields $A_1$ and $A_2$, 
the action can be written as 
\begin{equation}
\frac{ik}{4\pi}\int A\wedge d\tilde{A}-\int d^3x\,|\partial_\mu Y^I-iA_\mu Y^I|^2+\cdots. 
   \label{TFT+QFT}
\end{equation}
This action defines a theory of matters $Y^I$ coupled to a $\mathbb{Z}_k$ gauge theory constructed in terms of $A$ and $\tilde{A}$ \cite{Kapustin:2014gua}. 
This residual $\mathbb{Z}_k$ gauge symmetry makes the moduli space to be an orbifold. 
Since the superpotential vanishes for the case $k_1=-k_2$, the moduli space is $\mathbb{C}^4/\mathbb{Z}_k$ for ABJM theory. 

On the other hand, if $k_1+k_2$ is nonzero, then there is no choice of $A$ and $\tilde{A}$ for which the action becomes of the form (\ref{TFT+QFT}). 
An alternative choice gives 
\begin{equation}
\frac{ik_1k_2(k_1+k_2)}{4\pi}\int A\wedge dA+\frac{ik_1k_2}{4\pi}\int \tilde{A}\wedge d\tilde{A}-\int d^3x\,|\partial_\mu Y^I-i(k_1+k_2)AY^I|^2+\cdots. 
\end{equation}
Here, the gauge field $\tilde{A}$ simply decouples. 
Since the gauge group is originally ${\rm U}(1)\times{\rm U}(1)$, there are two sets of F-term conditions and D-term conditions. 
In the theory under consideration, these two sets of conditions are identical to each other. 
Therefore, the above theory can be regarded as a $U(1)$ theory coupled to three hypermultiplets, and as shown in \cite{Gukov:1999ya}, the resulting moduli space is $T^*\mathbb{P}^2$, without orbifolding. 

If the gravity duals for theories in ${\rm CSM}(3)$ in the planar limit are really eleven-dimensional, then the dictionary between the AdS$_4$ gravity and CFT$_3$ should be quite different from the one for ABJM theory. 
Then, the behavior of observables in the large 't~Hooft coupling limit might be quite different. 
For example, it is known that the dictionary for a CFT$_3$ whose gravity dual is a massive Type IIA theory is quite different from those for ABJM theory \cite{Aharony:2010af}. 
Due to this, the Wilson loop behaves as \cite{Suyama:2011yz} 
\begin{equation}
\langle W\rangle\ \sim\ \exp\left( ct^{\frac13} \right)
\end{equation}
for a constant $c$. 
The free energy scales as $N^2$ in the planar limit. 
The leading term of the free energy in the large 't~Hooft coupling limit is multiplied by a power of $t$ which reproduces the scaling in the M-theory limit \cite{Aharony:2010af}
\begin{equation}
F\ \sim\ c'N^\frac53. 
\end{equation}

It seems that more detailed investigations are necessary to understand the issue of gravity duals for theories in ${\rm CSM}(3)$. 

\vspace{1cm}

\section{Spectral curves} \label{section-curve}

\vspace{5mm}

In this section, we show that a difference among theories in ${\rm CSM}(n)$ can be found in the structure of planar spectral curves obtained from the resolvent $zv'(z)$. 
This observation can be extended to more general Chern-Simons-matter theories. 
The structure turns out to be governed by an associated Kac-Moody algebra specified by the matter representation of the theory. 
The reader may consult \cite{Carter} for results on Kac-Moody algebras mentioned below. 
The spectral curves of the theories in ${\rm CSM}(2)$ were discussed in \cite{Itoyama:2016imr}. 

\vspace{5mm}

\subsection{The theories in CSM($n$)} \label{curve for 2-node} \label{2-node curve}

\vspace{5mm}

It was shown in \cite{Suyama:2016nap} that the resolvent $v(z)$ satisfies the following vector equations: 
\begin{eqnarray}
(2\kappa_1,0) &=& x_1v'(x_1^+)-x_1v'(x_1^-)M_1, 
   \label{v-1} \\
(0,2\kappa_1) &=& x_2v'(x_2^+)-x_2v'(x_2^-)M_2, 
   \label{v-2}
\end{eqnarray}
where $x_\alpha\in I_\alpha$, and $x_\alpha^+$ ($x_\alpha^-$) is a point slightly above (below) $x_\alpha$. 
The matrices $M_\alpha$ are defined as 
\begin{equation}
M_1\ :=\ \left[
\begin{array}{cc}
-1 & 0 \\
n & 1
\end{array}
\right], \hspace{1cm} M_2\ :=\ \left[
\begin{array}{cc}
1 & n \\
0 & -1
\end{array}
\right]. 
\end{equation}
In many cases, the left-hand sides of (\ref{v-1})(\ref{v-2}) can be eliminated. 
This can be done by introducing $\omega(z)$ such that 
\begin{equation}
zv'(z)\ =\ c+\omega(z)
\end{equation}
is satisfied, where the constant row vector $c$ satisfies  
\begin{equation}
(2\kappa_1,2\kappa_2)\ =\ cA, \hspace{1cm} A\ :=\ \left[
\begin{array}{cc}
2 & -n \\
-n & 2
\end{array}
\right]. 
   \label{condition}
\end{equation}
The solution exists as long as $n\ne2$. 
Even in the case $n=2$, there exists a solution of this equation if and only if 
\begin{equation}
\kappa_1+\kappa_2\ =\ 0
\end{equation}
is satisfied. 
In the following, we assume that the equation (\ref{condition}) has a solution. 
Then, $\omega(z)$ satisfies 
\begin{equation}
\omega(x_\alpha^+)\ =\ \omega(x_\alpha^-)M_\alpha. 
   \label{homogeneous}
\end{equation}

\vspace{5mm}

It was shown in \cite{Suyama:2016nap} that, for the case $n=2$, the vector equations (\ref{homogeneous}) imply a simple scalar equation 
\begin{equation}
\omega_{\psi_2}(x_\alpha^+)\ =\ -\omega_{\psi_2}(x_\alpha^-), 
   \label{n=2}
\end{equation}
where 
\begin{equation}
\omega_\psi(z)\ :=\ \omega(z)\cdot\psi, \hspace{1cm} \psi_2\ :=\ \left[
\begin{array}{c}
1 \\
-1
\end{array}
\right]. 
\end{equation}
Therefore, $\omega_{\psi_2}(z)$ defines a two-sheeted covering of $\mathbb{C}$, that is, a torus. 

For the case $n=1$, one can check that the following equations are satisfied: 
\begin{equation}
\begin{array}{rclcrcl}
\omega_{\psi_1}(x_1^+) & = & \omega_{M_1\psi_1}(x_1^-), &\hspace{1cm}& \omega_{\psi_1}(x_2^+) & = & \omega_{\psi_1}(x_2^-), \\ [2mm]
\omega_{M_2M_1\psi_1}(x_1^+) & = & \omega_{M_2M_1\psi_1}(x_1^-), &\hspace{1cm}& 
\omega_{M_1\psi_1}(x_2^+) & = & \omega_{M_2M_1\psi_1}(x_2^-), 
\end{array}
\end{equation}
where 
\begin{equation}
\psi_1\ :=\ \left[
\begin{array}{c}
1 \\
0
\end{array}
\right]. 
\end{equation}
The above equations imply that the three scalar functions $\omega_{\psi_1}(z), \omega_{M_1\psi_1}(z), \omega_{M_2M_1\psi}(z)$ define a three-sheeted covering of $\mathbb{C}$. 
After a compactification, the resulting curve is topologically a sphere. 
Note that these scalar functions are given in terms of the components of the resolvent $v(z)$ as 
\begin{eqnarray}
\omega_{\psi_1}(z) &=& zv_1'(z)-c_1, \\
\omega_{M_1\psi_1}(z) &=& zv_2'(z)-zv_1'(z)+c_1-c_2, \\
\omega_{M_2M_1\psi_1}(z) &=& zv_2'(z)-c_2. 
\end{eqnarray}
Similar spectral curves appeared in a different context \cite{Dijkgraaf:2002vw}\cite{Itoyama:2009sc}. 

For the remaining cases $n\ge3$, it can be shown that $\omega_\psi(z)$ do not define a finite-sheeted covering of $\mathbb{C}$ for any choice of $\psi$. 
A proof is given in Appendix \ref{infinite sheets}. 
The best one can find is a finite covering with a twist, defined by the following equations: 
\begin{eqnarray}
\omega_{\psi_n}(x_1^+)\ =\ -\omega_{\psi_n'}(x_1^-), &\hspace{1cm}& \omega_{\psi_n}(x_2^+)\ =\ -e^{2\pi i\nu}\omega_{\psi_n'}(x_2^-), 
   \label{twisted covering}
\end{eqnarray}
where 
\begin{equation}
\psi_n\ :=\ \left[
\begin{array}{c}
1 \\
-e^{\pi i\nu}
\end{array}
\right], \hspace{1cm} \psi_n'\ =\ \left[
\begin{array}{c}
1 \\
-e^{-\pi i\nu}
\end{array}
\right]. 
\end{equation}
These equations are reduced to (\ref{n=2}) when $\nu=0$, or in other words, $n=2$. 

In summary, we have found that the spectral curve defined in terms of $\omega(z)$ is a Riemann surface of genus $n-1$ for the cases $n=1,2$, and otherwise, the curve is an infinite-covering of $\mathbb{C}$. 

There is also a difference between the case $n=1$ and $n=2$. 
For the case $n=1$, it can be shown that both $v_1'(z)$ and $v_2'(z)$ are algebraic functions, while for the case $n=2$, only the combination $v_1'(z)-v_2'(z)$ is an algebraic function. 

It is curious to notice that the matrix $A$ in (\ref{condition}) is a generalized Cartan matrix. 
The corresponding Kac-Moody algebra is ${\rm su}(3)$ for $n=1$, affine ${\rm su}(2)$ for $n=2$ and an algebra of indefinite type for $n\ge3$. 

\vspace{5mm}

\subsection{General quiver-type theories}

\vspace{5mm}

The structure of the spectral curves for theories in ${\rm CSM}(n)$ observed in the previous subsection can also be found in more general Chern-Simons-matter theories. 
Consider a Chern-Simons-matter theory with the gauge group $\prod_{a=1}^{n_g}{\rm U}(N_a)_{k_a}$ coupled to bi-fundamental hypermultiplets discussed in \cite{Jafferis:2008qz}\cite{Suyama:2013fua}. 
Let $n_{ab}$ be the number of the hypermultiplets coupled to ${\rm U}(N_a)\times {\rm U}(N_b)$ factor. 
Note that $n_{ab}$ is a symmetric matrix. 

Assume that the Chern-Simons levels $k_a$ are chosen such that 
\begin{equation}
2(\kappa_1,\cdots,\kappa_{n_g})\ =\ cA, \hspace{1cm} A_{bc}\ :=\ 2\delta_{bc}-n_{bc}
\end{equation}
has a solution. 
Again, the matrix $A$ can be regarded as the generalized Cartan matrix of a Kac-Moody algebra $\mathfrak{g}_A$. 
By construction, $A$ is always symmetric. 

As in the previous subsection, the planar analysis of the Chern-Simons-matter theory is reduced to solving the following vector equation \cite{Suyama:2013fua}: 
\begin{equation}
\omega(x_a^+)\ =\ \omega(x_a^-)M_a, 
   \label{gluing}
\end{equation}
where 
\begin{equation}
\omega(z)\ :=\ (\omega_1(z),\cdots, \omega_{n_g}(z))
\end{equation}
is a row-vector-valued function, and the matrices $M_a$ are defined as 
\begin{equation}
(M_a)_{bc}\ :=\ \delta_{bc}-\delta_{ac}A_{bc}, 
\end{equation}
where the repeated indices are not summed. 
It can be shown that $(M_a)^2=I$ is satisfied for any $a$. 

Let $W$ be a group generated by these $M_a$. 
To construct a covering of $\mathbb{C}$, we choose a column vector $\psi_0\in\mathbb{C}^{n_g}$ and consider the $W$-orbit: 
\begin{equation}
W\psi_0\ :=\ \{ w\psi_0\ |\ w\in W\ \}. 
\end{equation}
Define scalar functions $\omega_\psi(z):=\omega(z)\cdot\psi$ for each $\psi\in W\psi_0$. 
Then, the equations (\ref{gluing}) imply 
\begin{equation}
\omega_\psi(x_a^+)\ =\ \omega_{M_a\psi}(x_a^-), 
\end{equation}
which define a covering of $\mathbb{C}$. 
If $W\psi_0$ is a finite set, then the covering gives a Riemann surface of a finite genus. 

\vspace{5mm}

The following observation will be helpful for choosing an appropriate vector $\psi_0$ which defines a simple spectral curve. 
Consider the following vectors: 
\begin{equation}
(\alpha_a)_b\ :=\ A_{ab}. 
\end{equation}
The action of $M_a$ on $\alpha_b$ is 
\begin{equation}
M_a\alpha_b\ =\ \alpha_b-A_{ab}\alpha_a. 
\end{equation}
This indicates that $M_a$ act on $\alpha_b$ as if $\alpha_b$ are fundamental roots of $\mathfrak{g}_A$, and $M_a$ are the fundamental reflections of $\mathfrak{g}_A$. 
Then, $W$ gives a representation of the Weyl group of $\mathfrak{g}_A$. 

If the matrix $A$ is non-degenerate, then $\alpha_a$ indeed define fundamental roots in a root system $\Phi$ of $\mathfrak{g}_A$, and $M_a$ define a faithful representation of the Weyl group of $\mathfrak{g}_A$ on $\mathbb{C}^{n_g}$. 
Choosing $\psi_0$ as one of the roots $\alpha\in\Phi$, the $W$-orbit $W\psi_0$ is a subset of $\Phi$. 

\vspace{5mm}

Suppose that $\mathfrak{g}_A$ is of finite type, that is, $\mathfrak{g}_A$ is a finite-dimensional Lie algebra.  Then, $W\psi_0$ is a finite set since $\Phi$ is a finite set, and the covering defined above gives a Riemann surface of a finite genus. 

In fact, there is a simpler choice of $\psi_0$. 
Recall that the Weyl group also acts on the set of weights in an irreducible representation. 
If $\psi_0$ is chosen to be the highest weight of the smallest representation, then the resulting Riemann surface is the simplest possible one. 
The covering for the case $n=1$ given in the previous subsection is of this kind. 

\vspace{5mm}

Next, suppose that $\mathfrak{g}_A$ is of affine type. 
In this case, the rank of $A$ is $n_g-1$. 
This implies that $\alpha_a$ are not linearly independent, and therefore, $\alpha_a$ cannot define fundamental roots in a root system $\Phi$ of $\mathfrak{g}_A$. 
It is known that $\Phi$ can be realized in $\mathbb{C}^{n_g+1}$. 

The root system $\Phi$ of $\mathfrak{g}_A$ can be described explicitly as follows. 
Let $\mathfrak{g}_A^0$ be a finite-dimensional Lie algebra associated to  $\mathfrak{g}_A$, and let $\Phi^0$ be a root system of $\mathfrak{g}_A^0$. 
Then $\Phi$ is given as 
\begin{equation}
\Phi\ =\ \{ \alpha^0+r\delta \ |\ \alpha^0\in\Phi^0, r\in\mathbb{Z}\}, 
\end{equation}
where $\delta$ is defined as 
\begin{equation}
\delta\ :=\ \sum_{a=1}^{n_g}m_a\tilde{\alpha}_a
\end{equation}
in terms of the fundamental roots $\tilde{\alpha}_a\in\Phi$ of $\mathfrak{g}_A$ and integers $m_a$ satisfying 
\begin{equation}
\sum_{b=1}^{n_g}A_{ab}m_b\ =\ 0, \hspace{1cm} (m_1,m_2,\cdots,m_{n_g})\ =\ 1. 
\end{equation}

It is known that the one-dimensional subspace $L\subset\mathbb{C}^{n_g+1}$ spanned by $\delta$ is invariant under the action of the Weyl group of $\mathfrak{g}_A$. 
Then, the Weyl group also acts on the quotient space $\mathbb{C}^{n_g+1}/L$. 
There exists an isomorphism of vector spaces 
\begin{equation}
\pi\ :\ \mathbb{C}^{n_g+1}/L\ \to\ \mathbb{C}^{n_g}, \hspace{1cm} \pi(\tilde{\alpha}_a)\ =\ \alpha_a. 
\end{equation}
This is well-defined since 
\begin{equation}
\pi(\delta)_b\ =\ \sum_{a=1}^{n_g}m_a(\alpha_a)_b\ =\ \sum_{a=1}^{n_g}m_aA_{ab}\ =\ 0
\end{equation}
is satisfied. 
By this isomorphism, the matrices $M_a$ can be identified with the fundamental reflections of the Weyl group of $\mathfrak{g}_A$ acting on $ \mathbb{C}^{n_g+1}/L$. 
In this quotient space, $\Phi$ becomes equivalent to $\Phi^0$ which is a finite set. 
If $\psi_0$ is chosen to be one of $\alpha_a$, then the $W$-orbit $W\psi_0$ is again finite. 
The covering for the case $n=2$ given in the previous subsection is of this kind. 

\vspace{5mm}

If $\mathfrak{g}_A$ is of indefinite type, there does not seem to exist a choice of $\psi_0$ which gives us a Riemann surface of a finite genus as a spectral curve. 
It would be interesting if it would be possible to find a ``twisted'' covering like the one given in the previous subsection for the cases $n\ge3$. 
The existence of such coverings would open the possibility to determine the resolvent explicitly. 

\vspace{1cm}

\section{Discussion} \label{section-discuss}

\vspace{5mm}

We have discussed the large 't~Hooft coupling limit of theories in ${\rm CSM}(n\ge3)$. 
We have found that all the observables which can be calculated from the planar resolvent have finite limit in the large 't~Hooft coupling limit. 
As an example, the vevs of BPS Wilson loops are finite in the limit, quite different from the behavior observed in  ABJM theory. 
A proposal for the gravity dual of a theory in ${\rm CSM}(3)$ was revisited. 
In addition, we have found the structure of spectral curves of the theories which depends on $n$. 
The structure is governed by a Kac-Moody algebra associated to the matter representation of each Chern-Simons-matter theory. 

It is curious that the behavior of observables of theories in ${\rm CSM}(3)$ is quite different from the one observed in ABJM theory, although they seem to share many properties, like the dimension of the moduli space which would suggest the existence of a gravity dual. 
We briefly pointed out that the difference could be consistent with the observation that the dual gravity background would behave differently in the planar limit. 
To know more about the possible gravity duals of theories in ${\rm CSM}(3)$, we need to know more detailed properties of the CFT$_3$ side. 
One issue to be clarified is the consistency of the claim that the dual gravity background is AdS$_4\times N^{0,1,0}$ with \cite{Gaiotto:2009mv} since the Chern-Simons levels are chosen such that $k_1+k_2\ne0$ is satisfied. 

The observation that Kac-Moody algebras may play important role in Chern-Simons-matter matrix models seems to be quite interesting. 
The relation of Kac-Moody algebras and Chern-Simons-matter theories would imply a classification of the latter in terms of the classification of the former. 
This also suggests that detailed knowledge of Kac-Moody algebras of indefinite type would give us a crew to investigate various Chern-Simons-matter theories. 
For example, if one could find a representation of a Kac-Moody algebra of indefinite type which may give the twisted covering like (\ref{twisted covering}) given in subsection \ref{2-node curve}, then the planar resolvent of the corresponding Chern-Simons-matter theory could be obtained explicitly in terms of the theta functions on a higher-genus Riemann surface. 

\vspace{2cm}

{\bf \Large Acknowledgements}

\vspace{5mm}

We would like to thank H. Itoyama, Y. Matsuo, T. Okazaki, T. Oota, R. Yoshioka for valuable discussions. 
This work was supported in part by the Grant-in-Aid for Scientific research, No 16H06490 and Fujukai Foundation.

\appendix

\vspace{2cm}

\section{Details on the planar resolvent} \label{details}

\vspace{5mm}

In this appendix, we review some details on the planar resolvent obtained in \cite{Suyama:2016nap}. 

The derivative of the planar resolvent can be written as 
\begin{equation}
zv'(z)\ =\ c+\frac1{s(z)}f(z), 
\end{equation}
where $s(z)$ is a square-root function defined in (\ref{v to f}), and $c$ is a constant vector satisfying (\ref{vector c}). 
Explicitly, 
\begin{equation}
c_1\ =\ -\frac{4\kappa_1+2n\kappa_2}{n^2-4}, \hspace{1cm} c_2\ =\ -\frac{2n\kappa_1+4\kappa_2}{n^2-4}
\end{equation}
for $n\ne2$. 
The row-vector-valued function $f(z)$ is given in terms of a scalar-valued function $\tilde{f}(u)$ as in (\ref{S}). 

One can check that a function $G(u)$ defined as 
\begin{equation}
G(u)\ :=\ \frac{\vartheta_1(u-u_\nu)\vartheta_1(u-u_\nu+\frac12)}{\vartheta_1(u-u_\infty)\vartheta_1(u+u_\infty)}
\end{equation}
satisfies the conditions (\ref{conditions tilde{f}}), provided that the parameters are defined as 
\begin{equation}
u_0\ :=\ u(0), \hspace{1cm} u_\infty\ :=\ u(\infty), \hspace{1cm} u_\nu\ :=\ \frac12\nu+\frac14. 
\end{equation}
In terms of $G(u)$, the function $\tilde{f}(u)$ is given as 
\begin{equation}
\tilde{f}(u)\ =\ g(u)G(u)
\end{equation}
where $g(u)$ is an elliptic function. 
The function $g(u)$ is then determined by requiring that the derivative $zv'(z)$ has the right analytic structure. 
The explicit form of $g(u)$ is 
\begin{equation}
g(u)\ =\ r_1(1+z(u))+r_3(g_3(u)+g_4(u)), 
\end{equation}
where $z(u)$ is defined in (\ref{z(u)}), and 
\begin{equation}
g_3(u)\ :=\ \frac{\vartheta_1(u-u_0)\vartheta_1(u-u_\nu+\frac12)}{\vartheta_1(u-u_\infty)\vartheta_1(u-u_\nu)}, \hspace{1cm} g_4(u)\ :=\ -\frac{\vartheta_1(u+u_0)\vartheta_1(u-u_\nu)}{\vartheta_1(u+u_\infty)\vartheta_1(u-u_\nu+\frac12)}. 
\end{equation}
The coefficients $r_1,r_3$ are given as 
\begin{eqnarray}
r_1 &=& \frac1{g_3(-u_0)-g_4(u_0)}\left[ \tilde{c}_1\frac{g_3(-u_0)}{G(u_0)}-\tilde{c}_2\frac{g_4(u_0)}{G(-u_0)} \right], \\ [2mm]
r_3 &=& \frac1{g_3(-u_0)-g_4(u_0)}\left[ -\frac{\tilde{c}_1}{G(u_0)}+\frac{\tilde{c}_2}{G(-u_0)} \right], 
\end{eqnarray}
where $\tilde{c}:=(\tilde{c}_1,\tilde{c}_2)$ is defined as $\tilde{c}:=cS$. 
Note that the equality 
\begin{equation}
g_3(-u_0)-g_4(u_0)\ =\ g_3(-u_0)\left( 1+z(u_\nu) \right)
\end{equation}
holds. 

\vspace{1cm}

\section{Infinite number of sheets for $n\ge3$} \label{infinite sheets} 

\vspace{5mm}

Let $W$ be a group generated by $M_1$ and $M_2$. 
As explained in subsection \ref{curve for 2-node}, one can construct a spectral curve from the equations (\ref{homogeneous}) in terms of the scalar functions 
\begin{equation}
\omega_{\psi}(z), \hspace{10mm} \psi\ \in\ W\psi_0
\end{equation}
for a choice of $\psi_0\in\mathbb{C}^2$. 

Suppose that $W\psi_0$ is finite for a suitable $\psi_0$. 
Then, there exists an element $w\in W$ such that 
\begin{equation}
w\psi_0\ =\ \psi_0
   \label{finite}
\end{equation}
holds. 
The element $w$ has one of the following forms: 
\begin{equation}
(M_1M_2)^m, \hspace{5mm} (M_2M_1)^m, \hspace{5mm} M_1(M_2M_1)^m, \hspace{5mm} M_2(M_1M_2)^m
\end{equation}
for a non-negative integer $m$. 
One may choose a basis of $\mathbb{C}^2$ such that $M_1$ and $M_2$ can be written as 
\begin{equation}
M_1\ =\ \left[ 
\begin{array}{cc}
0 & -1 \\
-1 & 0
\end{array}
\right], \hspace{1cm} M_2\ =\ \left[ 
\begin{array}{cc}
0 & -e^{-2\pi i\nu} \\
-e^{2\pi i\nu} & 0
\end{array}
\right]. 
   \label{rep of M}
\end{equation}
For $n\ge3$, the parameter $\nu$ is purely imaginary. 
We choose ${\rm Im}(\nu)>0$. 

From (\ref{rep of M}), one obtains 
\begin{equation}
(M_1M_2)^m\ =\ \left[ 
\begin{array}{cc}
e^{2\pi im\nu} & \\
0 & e^{-2\pi im\nu}
\end{array}
\right]. 
\end{equation}
Apparently, if $w$ is of this form, then the condition (\ref{finite}) cannot be satisfied for any $\psi_0$. 
By the same reasoning, $w$ is not of the form $(M_2M_1)^m$. 

Now, suppose $w=M_2(M_1M_2)^m$. 
The explicit form of this matrix is 
\begin{equation}
\left[
\begin{array}{cc}
0 & -e^{2\pi i(m+1)\nu} \\
-e^{2\pi i(m+1)\nu} & 0
\end{array}
\right]. 
\end{equation}
Then, $\psi_0$ is determined to be 
\begin{equation}
\psi_0\ =\ \left[
\begin{array}{c}
1 \\
-e^{2\pi i(m+1)\nu}
\end{array}
\right]
\end{equation}
up to an overall factor. 
This $\psi_0$ gives 
\begin{eqnarray}
(M_1M_2)^l\psi_0 &=& \left[
\begin{array}{c}
e^{2\pi il\nu} \\
-e^{2\pi i(m+1-l)\nu}
\end{array}
\right], \\
M_2(M_1M_2)^l\psi_0 &=& \left[
\begin{array}{c}
e^{2\pi i(m-l)\nu} \\
-e^{2\pi i(l+1)\mu}
\end{array}
\right]
\end{eqnarray}
for $0\le l\le m$. 
If $W\psi_0$ is finite, then $(M_2M_1)^l\psi$ for a positive integer $l$ must have one of the above forms. 
However, this is impossible since its form is 
\begin{equation}
(M_2M_1)^l\psi\ =\ \left[
\begin{array}{c}
e^{-2\pi il\nu} \\
-e^{2\pi i(m+1+l)\nu}
\end{array}
\right]. 
\end{equation}
Therefore, $w$ is not of the form $M_2(M_1M_2)^m$. 
The same argument excludes the possibility for $w$ to be of the form $M_1(M_2M_1)^m$. 
It is concluded that an element $w\in W$ satisfying (\ref{finite}) does not exist, implying that $W\psi_0$ is infinite for any choice of $\psi_0$.

\end{document}